\documentstyle[12pt]{article}
\advance\oddsidemargin by -\textwidth
\advance\oddsidemargin by -1in
\evensidemargin=\oddsidemargin

\makeatletter
\def\@maketitle{\newpage
 \null
 {\normalsize \tt \begin{flushright}
  \begin{tabular}[t]{l} \@date
  \end{tabular}
 \end{flushright}}
 \begin{center}
 \vskip 2em
 {\LARGE \@title \par} \vskip 1.5em {\large \lineskip .5em
 \begin{tabular}[t]{c}\@author
 \end{tabular}\par}
 \end{center}
 \par
 \vskip 1.5em}
\makeatother
\newcommand{\tMCpr}[1]{{\hat \alpha}^{#1}_\perp}   
\newcommand{\tMClr}[1]{{\hat \alpha}^{#1}_\parallel}
\newcommand{\cV}{{\cal V}} 
\newcommand{\tr}{\mbox{tr}}
\title{
  Formulations of spin 1 resonances \\
  in the chiral lagrangian}
\author{
  {\sc Masaharu Tanabashi}\thanks{
    E-mail address: {\tt tanabash@theory.kek.jp}}
 \\
  {\it National Laboratory for High Energy Physics (KEK)} \\
  {\it Tsukuba, Ibaraki 305, Japan}
}
\date{
  KEK-TH-438 \\
  KEK-preprint 95-156 \\
  November 1995
}
\begin{document}
\maketitle
\begin{abstract}
Equivalence of the hidden local symmetry formulation with non-minimal
interactions and the anti-symmetric tensor field method of
$\rho$ and $a_1$ mesons in the chiral lagrangian is shown by using
the auxiliary field method.
Violation of the KSRF I relation, which becomes important in the
application of chiral lagrangian to {\em non QCD-like}
technicolor models can be parametrized by non-minimal coupling in
the hidden local symmetry formalism keeping low energy theorem of
hidden local symmetry.
We also obtain explicit correspondence of parameters in both formulations.
\end{abstract}

The vector meson plays important roles in the chiral lagrangian of the
spontaneous chiral symmetry breaking in QCD.\@
The observed sizes of chiral coefficients $L_{1,2,3,9,10}$
in one loop chiral perturbation theory\cite{kn:GL84} at ${\cal
O}(E^4)$ are saturated by the vector meson contribution\cite{kn:EGPR89}.

In the application of the chiral lagrangian to the strongly interacting
Higgs sector, the chiral coefficients $L_{10}$ and $L_9$
correspond\cite{kn:Ho91} to the Peskin--Takeuchi $S$
parameter\cite{kn:PT90} and the anomalous triple gauge boson interactions
$\Delta \kappa_{\gamma,Z}$\cite{kn:HPZH}.
Thus, the measurements of these parameters give severe mass bounds on
the techni-$\rho$ of the QCD-like technicolor model.

It should be emphasized, however, that the naive QCD-like technicolor
models already suffer from the serious disease of excess of
flavor changing neutral currents.
We thus  need to consider {\em non QCD-like} technicolor models,
e.g., the walking technicolor model\cite{kn:Walk} and the technicolor
model with an elementary scalar\cite{kn:Si89}, etc..
Unlike the naive QCD-like technicolor model, these non QCD-like models
are considered to have relatively hard high energy behavior of the
Nambu-Goldstone boson form factor due to the large anomalous
dimension or the appearance of the elementary scalar.

Many formulations to incorporate the $\rho$ meson into the chiral
lagrangian have been proposed.
One of the most famous formulations was proposed by Bando, Kugo,
Uehara, Yamawaki and Yanagida (BKUYY), in which the $\rho$ meson is
treated as a gauge field of ``hidden local symmetry'' in the chiral
lagrangian\cite{kn:BKUYY,kn:BKY88}.
In the absence of the external gauge fields ($W$ and photon),
the BKUYY formulation has two free parameters ($a$, $g$) in addition to
$f_\pi$.
It thus describes the most general amplitude of the $\rho\pi\pi$
coupling and the mass of the $\rho$-meson.
The BKUYY formulation, however, fixes the amplitude of the mixing of
the external gauge field and the $\rho$ meson field, leading to the
KSRF\cite{kn:KS66} I relation\cite{kn:BKY85n,kn:BKY85,kn:HY92}.

The ``vector limit'' model of the $\rho$ meson\cite{kn:Ge89} can be
considered as a special case of this model ($a=1$).
An one loop calculation is performed based on the BKUYY
formulation\cite{kn:HY92,kn:Ta93}.
The technicolored version of this model is known as the BESS
model\cite{kn:BESS}.
Despite the great success of this model in QCD and QCD-like
technicolor, the BKUYY formulation is not appropriate for the analysis
of the non QCD-like technicolor model as it stands, since the KSRF I
relation in QCD is a manifestation of the soft high
energy behavior of the pion form factor.


Yet another popular formulation of the $\rho$ meson was proposed by
Gasser and
Leutwyler\cite{kn:GL84,kn:EGPR89}, in which the $\rho$ meson is
represented by an anti-symmetric tensor field.
In the absence of external gauge fields, this formulation has two
parameters ($G_V$, $M_V$) corresponding to the
$\rho\pi\pi$ coupling and the mass of the $\rho$ meson in addition to
$f_\pi$.
This model is equivalent to the usual vector field
formulation including hidden local symmetry formulation in the
absence of the  external gauge field in a Hamiltonian
language\cite{kn:Kala93}.

Unlike the hidden local symmetry formalism,
the $\rho$-photon mixing amplitude is left to be a free parameter
$F_V$ in this model.
Although this extended parameter space is suited for the
analysis of non QCD-like technicolor model, actual
calculations e.g. one loop chiral logarithms, are difficult
in this model due to its complicated Feynman rules.


It has been checked that the
vector meson contribution to the low energy chiral coefficients
$L_{1,2,3,9,10}$ are independent of the
choice of the formulations for the case of QCD\cite{kn:EGLPR89}
where the KSRF I relation is known to be satisfied phenomenologically.
%
However, the difference of the formulations becomes serious
in the {\em non QCD-like} technicolor models.
Actually the BKUYY formulation predicts $L_{10}=-L_9$ as a consequence
of the KSRF I relation, which leads to
serious cancellations in $\Delta \kappa_{\gamma,Z}$\cite{kn:Ho91},
while the anti-symmetric tensor method does not give such a prediction.


The aim of this paper is to give a simple method to study the
relation between both formulations by using an auxiliary field method.
We find that the anti-symmetric tensor method becomes equivalent to the
hidden local symmetry formalism after adding several ${\cal O}(E^4)$
parameters in the hidden local lagrangian.

For simplicity, we first study the effective lagrangian of the $\pi$
and the $\rho$ meson without including the $a_1$ meson. The effect of
the $a_1$ meson will be discussed later.

Let us start with the conventional chiral lagrangian of
$SU(N)_L\times SU(N)_R/SU(N)_V$ symmetry so as to fix our notations:
\begin{equation}
  {\cal L} = \frac{f_\pi^2}{4} \mbox{tr}((D^\mu U)^\dagger (D_\mu U)),
  \qquad
  U = \exp(2i\frac{\pi^a T^a}{f_\pi}),
\end{equation}
with $T^a$ the $SU(N)$ generator.
Here the chiral field $U$ transforms as $U\rightarrow g_L U
g_R^\dagger$ under $SU(N)_L \times SU(N)_R$.
The covariant derivative $D_\mu$ is given by
\begin{displaymath}
  D_\mu U = \partial_\mu U - i {\cal L}_\mu U + i U {\cal R}_\mu.
\end{displaymath}
Here, we introduced the external gauge fields ${\cal L}_\mu$ and
${\cal R}_\mu$ corresponding to the chiral symmetry
$SU(N)_L \times SU(N)_R$.

In ref\cite{kn:GL84,kn:EGPR89} a $\rho$ meson field is introduced as an
anti-symmetric tensor field ${\bf V}_{\mu\nu}$ with the matter--type
transformation properties:
\begin{eqnarray}
  {\cal L}_{\rm AST}
    &=& \frac{f_\pi^2}{4} \mbox{tr}((D^\mu U)^\dagger (D_\mu U))
       -\frac{1}{2} \mbox{tr}(\nabla^\lambda {\bf V}_{\lambda\mu}
                               \nabla_\nu {\bf V}^{\nu\mu})
        +\frac{M_V^2}{4} \mbox{tr}({\bf V}_{\mu\nu}{\bf V}^{\mu\nu})
    \nonumber\\
    & & +\frac{F_V}{\sqrt{2}}\tr({\bf V}^{\mu\nu} \hat{\cal V}_{\mu\nu})
        +\frac{G_V}{2\sqrt{2}}
           i \mbox{tr}({\bf V}^{\mu\nu}[u_\mu,u_\nu]).
\label{eq:AST}
\end{eqnarray}
We define $\xi$ as
\begin{equation}
   U = \xi \xi,
\label{eq:div1}
\end{equation}
and the covariant derivative $\nabla_\mu$ is given by
\begin{displaymath}
  \nabla^\lambda {\bf V}_{\lambda\mu}
    = \partial^\lambda {\bf V}_{\lambda\mu}
     + [ \Gamma^\lambda, {\bf V}_{\lambda\mu} ],
  \quad
  \Gamma_\mu = -\frac{1}{2} ( \partial_\mu \xi^\dagger \cdot \xi
                              +\partial_\mu \xi \cdot \xi^\dagger
                              +i\xi^\dagger {\cal L}_\mu \xi
                              +i\xi {\cal R}_\mu \xi^\dagger ).
\end{displaymath}
The chiral covariant one form $u_\mu$ and the external vector field
strength $\hat{\cal V}_{\mu\nu}$ are defined by
\begin{displaymath}
  u_\mu = i \xi^\dagger (D_\mu U) \xi^\dagger,
  \qquad
  \hat{\cal V}_{\mu\nu} = \frac{1}{2}
         ( \xi^\dagger{\cal L}_{\mu\nu}\xi
          +\xi{\cal R}_{\mu\nu}\xi^\dagger ).
\end{displaymath}

This model has four parameters $f_\pi, M_V, F_V, G_V$ corresponding to
the pion decay constant, the mass of the $\rho$ meson, $\rho$-$\gamma$
mixing and $\rho\pi\pi$ coupling, respectively.  The KSRF I relation
is expressed by the relation $F_V=2G_V$.

The space components of the anti-symmetric tensor field ${\bf
V}_{ij}=-{\bf V}_{ji}$ ($i,j=1,2,3$) do not have time derivatives and
thus they can be removed from the dynamics.
The time components ${\bf V}_{0i}=-{\bf V}_{i0}$ are the dynamical
degrees of freedom identified as the $\rho$ meson.

Bando, Kugo, Uehara, Yamawaki and Yanagida (BKUYY)\cite{kn:BKUYY}
noted that the decomposition (\ref{eq:div1}) to include matter fields
in the chiral lagrangian can be extended to the following form:
\begin{equation}
  U = \xi_L^\dagger \xi_R.
\label{eq:div2}
\end{equation}
The decomposition (\ref{eq:div2}) has an ambiguity in the
definition of $\xi_L$ and $\xi_R$.
BKUYY regarded this ambiguity as a symmetry (hidden local symmetry,
$H=SU(N)$)
\begin{equation}
  \xi_L \rightarrow h \xi_L g_L^\dagger, \qquad
  \xi_R \rightarrow h \xi_R g_R^\dagger, \qquad h\in H.
\end{equation}

By introducing the $\rho$ meson as a gauge field of the above hidden
local symmetry, BKUYY proposed a lagrangian:
\begin{equation}
  {\cal L}_{\rm BKUYY}
    =  f_\pi^2 \mbox{tr}(\hat\alpha_{\mu\perp} \hat\alpha^\mu_\perp)
     +af_\pi^2 \mbox{tr}(\hat\alpha_{\mu\parallel} \hat\alpha^\mu_\parallel)
     -\frac{1}{2g^2} \mbox{tr}(V_{\mu\nu} V^{\mu\nu}),
\label{eq:hls1}
\end{equation}
where $\hat\alpha_{\mu\perp}$ and $\hat\alpha_{\mu\parallel}$ are given by
\begin{displaymath}
  \hat\alpha_{\mu\perp}
  \equiv \frac{1}{2i} \left(
           D_\mu \xi_L \cdot \xi_L^\dagger
         - D_\mu \xi_R \cdot \xi_R^\dagger
        \right),
\qquad
  \hat\alpha_{\mu\parallel}
  \equiv \frac{1}{2i} \left(
           D_\mu \xi_L \cdot \xi_L^\dagger
         + D_\mu \xi_R \cdot \xi_R^\dagger
        \right),
\end{displaymath}
with the covariant derivative $D_\mu$ being
\begin{displaymath}
  D_\mu \xi_L = \partial_\mu \xi_L - i V_\mu \xi_L + i \xi_L {\cal L}_\mu,
  \qquad
  D_\mu \xi_R = \partial_\mu \xi_R - i V_\mu \xi_R + i \xi_R {\cal R}_\mu.
\end{displaymath}
The gauge field $V_\mu$ is identified with the $\rho$ meson.

This model is parametrized by 3 parameters ($f_\pi,a,g$), while it
explains 4 physical quantities (the pion decay constant, the
$\rho$-$\gamma$ mixing, the mass of the $\rho$ meson and the $\rho\pi\pi$
couplings).
This model thus has one prediction, corresponding to the KSRF I relation.
Actually the KSRF I at zero momentum is derived as a low
energy theorem of the hidden local symmetry.
However, it should be noted that the off-shell relation in the
effective field theory might be unphysical because it depends on the
definition of the effective fields.
Actually as we will show later, the physical on-shell KSRF I relation
can be violated by
adding higher derivative terms keeping manifest hidden local symmetry.

We note here that the hidden local symmetry can be
made manifest without introducing a corresponding gauge field.
Actually, by changing the definition of $\Gamma_\mu$, $u_\mu$
and $\hat{\cal V}_{\mu\nu}$:
\begin{eqnarray*}
  \Gamma_\mu &=& -\frac{1}{2} (
                    \partial_\mu \xi_L \cdot \xi_L^\dagger
                   +\partial_\mu \xi_R \cdot \xi_R^\dagger
                   +i\xi_L {\cal L}_\mu \xi_L^\dagger
                   +i\xi_R {\cal R}_\mu \xi_R^\dagger ),
  \\
  u_\mu &=& i \xi_L (D_\mu U) \xi_R^\dagger,
  \\
  \hat{\cal V}_{\mu\nu}
  &=& \frac{1}{2}
         ( \xi_L {\cal L}_{\mu\nu}\xi_L^\dagger
          +\xi_R {\cal R}_{\mu\nu}\xi_R^\dagger )
\end{eqnarray*}
the anti-symmetric tensor lagrangian (\ref{eq:AST}) becomes invariant
under the hidden local symmetry.
In this case $\Gamma_\mu$ plays the role of the gauge connection of
the hidden local symmetry.
We further obtain the following relations between the anti-symmetric tensor
and the BKUYY notations:
\begin{displaymath}
  \Gamma_\mu = -i (\hat\alpha_{\mu\parallel}+V_\mu), \qquad
  u_\mu = 2 \hat\alpha_{\mu\perp}.
\end{displaymath}

Now, we are ready to show the equivalence of both formulations.
We introduce an auxiliary field $V_\mu$ into the lagrangian
(\ref{eq:AST}) of the anti-symmetric tensor formalism.
The dynamics is not modified by adding an auxiliary field $V_\mu$:
\begin{equation}
  {\cal L}'_{\rm AST}
    = {\cal L}_{\rm AST}
     + \frac{\kappa^2}{2}
       \tr\left(
         (V_\mu -i\Gamma_\mu
           -\frac{1}{\kappa} \nabla^{\nu} {\bf V}_{\nu\mu})^2
       \right),
\label{eq:aux0}
\end{equation}
with $\kappa$ an arbitrary parameter.
The lagrangian (\ref{eq:aux0}) then reads:
\begin{eqnarray}
{\cal L}'_{\rm AST}
  &=& f_\pi^2 \mbox{tr}(\hat\alpha_{\mu\perp}\hat\alpha^{\mu}_{\perp})
     +\kappa
      \tr(\hat \alpha^\nu_\parallel \nabla^{\mu} {\bf V}_{\mu\nu})
     +\frac{\kappa^2}{2} \tr(\hat\alpha_{\mu\parallel}
                             \hat\alpha^\mu_\parallel)
     +\frac{M_V^2}{4} \tr({\bf V}_{\mu\nu} {\bf V}^{\mu\nu})
  \nonumber\\
  & & \quad
   +\frac{F_V}{\sqrt{2}} \tr({\bf V}_{\mu\nu} \hat{\cal V}^{\mu\nu})
   +\sqrt{2} G_V i\tr({\bf V}_{\mu\nu}
                      [ \hat\alpha^{\mu}_\perp, \hat\alpha^{\nu}_\perp ] ),
\label{eq:EGLPR2}
\end{eqnarray}
Performing a partial integral, we can remove the derivative term of the
anti-symmetric tensor field:
\begin{equation}
  \tr(\hat \alpha^\nu_\parallel \nabla^{\mu} {\bf V}_{\mu\nu})
  = -\frac{1}{2} \tr(( D^\mu \hat \alpha^\nu_\parallel
                        -D^\nu \hat \alpha^\mu_\parallel )
                       {\bf V}_{\mu\nu})
      + i \tr([\hat\alpha^\mu_\parallel, \hat\alpha^\nu_\parallel]
            {\bf V}_{\mu\nu}
        ),
\label{eq:partial}
\end{equation}
where we have defined the covariant derivative
\begin{math}
  D_\mu \hat\alpha_{\nu\parallel} \equiv
    \partial_\mu \hat\alpha_{\nu\parallel}
   -i[V_\mu, \hat\alpha_{\nu\parallel} ]
\end{math}.
We note here that the partial integral (\ref{eq:partial}) transfers
the dynamical degree of freedom from the anti-symmetric tensor field
${\bf V}_{\mu\nu}$ to the auxiliary field $V_\mu$.
Plugging an identity
\begin{displaymath}
  D_\mu\hat\alpha_{\nu\parallel}-D_\nu \hat\alpha_{\mu\parallel}
  = i[\hat\alpha_{\mu\parallel}, \hat\alpha_{\nu\parallel}]
   +i[\hat\alpha_{\mu\perp}, \hat\alpha_{\nu\perp}]
   +\hat{\cal V}_{\mu\nu} - V_{\mu\nu},
\end{displaymath}
with
\begin{math}
  V_{\mu\nu} \equiv \partial_\mu V_\nu - \partial_\nu V_\mu
                   -i[V_\mu, V_\nu]
\end{math}
into (\ref{eq:partial}), we find
\begin{displaymath}
  \tr(\hat \alpha^\nu_\parallel \nabla^{\mu} {\bf V}_{\mu\nu})
  = \frac{i}{2}\tr(
         [\hat\alpha^\mu_\parallel, \hat\alpha^\nu_\parallel] {\bf V}_{\mu\nu}
      )
     -\frac{i}{2}\tr(
         [\hat\alpha^\mu_\perp, \hat\alpha^\nu_\perp]{\bf V}_{\mu\nu}
      )
     -\frac{1}{2} \hat{\cal V}_{\mu\nu}
     +\frac{1}{2} V_{\mu\nu}.
\end{displaymath}
The lagrangian (\ref{eq:EGLPR2}) then reads
\begin{eqnarray}
  {\cal L}_{\rm AST}'
  &=&  f_\pi^2 \mbox{tr}(\hat\alpha_{\mu\perp}\hat\alpha^{\mu}_{\perp})
     + \frac{M_V^2}{4} \tr({\bf V}_{\mu\nu} {\bf V}^{\mu\nu})
  \nonumber\\
  & &+ \frac{i}{2} \kappa
       \tr({\bf V}^{\mu\nu} [ \hat\alpha_{\mu\parallel},
                               \hat\alpha_{\nu\parallel} ] )
     -i \left(\frac{\kappa}{2} - \sqrt{2} G_V \right)
       \tr({\bf V}^{\mu\nu} [ \hat\alpha_{\mu\perp},
                               \hat\alpha_{\nu\perp} ] )
  \nonumber\\
  & &+\frac{\kappa}{2} \tr({\bf V}^{\mu\nu} V_{\mu\nu})
     -\left( \frac{\kappa}{2} - \frac{F_V}{\sqrt{2}}\right)
        \tr({\bf V}^{\mu\nu} \hat {\cal V}_{\mu\nu})
     +\frac{\kappa^2}{2} \tr(\hat\alpha_{\mu\parallel}
                             \hat\alpha^\mu_{\parallel}).
\end{eqnarray}
It is easy to integrate out the anti-symmetric tensor field ${\bf
V}_{\mu\nu}$:
\begin{eqnarray}
  {\cal L}_{\rm AST}'
  &=& f_\pi^2 \mbox{tr}(\hat\alpha_{\mu\perp}\hat\alpha^\mu_{\perp})
     +\frac{\kappa^2}{2} \tr(\hat\alpha_{\mu\parallel}
                             \hat\alpha^\mu_{\parallel})
  \nonumber\\
  & & -\frac{1}{4M_V^2} \tr\left(
       \left(
         \kappa V_{\mu\nu}
        -(\kappa - \sqrt{2} F_V) \hat{\cal V}_{\mu\nu}
        +\kappa i[\hat\alpha_{\mu\parallel},\hat\alpha_{\nu\parallel}]
        -(\kappa -2\sqrt{2}G_V)
         i[\hat\alpha_{\mu\perp},\hat\alpha_{\nu\perp}]
       \right)^2 \right).
  \nonumber\\
  & &
\label{eq:EGLPR3}
\end{eqnarray}
The lagrangian (\ref{eq:EGLPR3}) is equivalent to
the BKUYY lagrangian with several extra terms:
\begin{eqnarray}
  {\cal L}_{\rm BKUYY}
    &=&  f_\pi^2 \mbox{tr}(\hat\alpha_{\mu\perp} \hat\alpha^\mu_\perp)
     +af_\pi^2 \mbox{tr}(\hat\alpha_{\mu\parallel} \hat\alpha^\mu_\parallel)
     -\frac{1}{2g^2} \mbox{tr}(V_{\mu\nu} V^{\mu\nu})
    \nonumber\\
  & &+z_1\tr(\hat\cV_{\mu\nu}\hat\cV^{\mu\nu})
     +z_3\tr(\hat\cV_{\mu\nu} V^{\mu\nu})
     +z_4 i\tr(V_{\mu\nu}\tMCpr{\mu}\tMCpr{\nu})
     +z_5 i\tr(V_{\mu\nu}\tMClr{\mu}\tMClr{\nu})
  \nonumber\\
  & &
     +z_6 i\tr(\hat\cV_{\mu\nu}\tMCpr{\mu}\tMCpr{\nu})
     +z_7 i\tr(\hat\cV_{\mu\nu}\tMClr{\mu}\tMClr{\nu})
     +\cdots,
\label{eq:BKUYYextra}
\end{eqnarray}
with
\vspace*{-1cm}
\begin{displaymath}
\begin{minipage}[t]{3in}
\begin{eqnarray*}
  a   &=& \frac{\kappa^2}{2f_\pi^2},
      \\
  z_1 &=& -\frac{(\kappa-\sqrt{2}F_V)^2}{4M_V^2},
      \\
  z_4 &=& \frac{\kappa(\kappa-2\sqrt{2}G_V)}{M_V^2},
      \\
  z_6 &=& -\frac{(\kappa-\sqrt{2}F_V)(\kappa-2\sqrt{2}G_V)}{M_V^2},
      \\
\end{eqnarray*}
\end{minipage}
\begin{minipage}[t]{3in}
\begin{eqnarray*}
  g   &=& \frac{\sqrt{2}M_V}{\kappa},
      \\
  z_3 &=& \frac{\kappa(\kappa-\sqrt{2}F_V)}{2M_V^2},
      \\
  z_5 &=& -\frac{\kappa^2}{M_V^2},
      \\
  z_7 &=& \frac{\kappa(\kappa-\sqrt{2}F_V)}{M_V^2},
\end{eqnarray*}
\end{minipage}
\end{displaymath}
where we have used the notation of Ref\cite{kn:Ta93} and $\cdots$ in
(\ref{eq:BKUYYextra}
stands for operators corresponding to four point vertices.

The above procedure, however, leaves the artificial coefficient
$\kappa$.  What is the meaning of $\kappa$, then?
Since we are dealing with an effective theory, there is an ambiguity in
the definition of effective fields.
The redefinition of the effective field does not change the physical
on-shell $S$ matrix, even though it modifies parameters in the effective
lagrangian.
The arbitrary parameter $\kappa$ corresponds to
this superficial parameter difference as we will show in the next paragraph.

Let us consider the redefinition of the $\rho$ meson field in the
hidden local symmetry formulation:
\begin{displaymath}
  V_\mu \rightarrow V_\mu + (1-K) \hat \alpha_{\mu\parallel}.
\end{displaymath}
This redefinition leads to:
\begin{eqnarray}
  V_{\mu\nu}
    &\rightarrow& K V_{\mu\nu} + (1-K) \hat {\cal V}_{\mu\nu}
       + K(1-K) i [\hat \alpha_{\mu\parallel},
                   \hat \alpha_{\nu\parallel}]
       + (1-K) i [\hat \alpha_{\mu\perp}, \hat \alpha_{\nu\perp}],
    \nonumber\\
  \hat\alpha_{\mu\parallel}
    &\rightarrow& K \hat\alpha_{\mu\parallel}.
\label{eq:redef}
\end{eqnarray}
Plugging (\ref{eq:redef}) into (\ref{eq:EGLPR3}), we find that the $\kappa$
dependence of (\ref{eq:EGLPR3}) appears only in the form $\kappa K$.
The arbitrary parameter $K$ thus actually corresponds to the ambiguity of
$\kappa$.

What is the most convenient choice of the $\rho$ meson field
definition, then?
One plausible choice is to define the $\rho$ meson field so as to
eliminate one of the non-minimal couplings $z_{1,3,4,5,6,7}$.
In the following, we choose a $\rho$ meson field definition in which
the kinetic $\rho$-$\gamma$ mixing $z_3$ is absent
($\kappa=\sqrt{2}F_V$).

This particular choice of the $\rho$ meson field definition resolves
the ambiguity of $\kappa$. We find the following relations:
\vspace*{-1cm}
\begin{equation}
\begin{minipage}[t]{3in}
\begin{eqnarray*}
  a   &=& \frac{F_V^2}{f_\pi^2},
      \\
  z_1 &=& 0,
      \\
  z_4 &=& \frac{2F_V}{M_V^2}(F_V-2G_V),
      \\
  z_6 &=& 0,
\end{eqnarray*}
\end{minipage}
\begin{minipage}[t]{3in}
\begin{eqnarray*}
  g   &=& \frac{M_V}{F_V},
      \\
  z_3 &=& 0,
      \\
  z_5 &=& -\frac{2F_V^2}{M_V^2},
      \\
  z_7 &=& 0.
\end{eqnarray*}
\end{minipage}
\end{equation}
Note here that the violation of the KSRF I relation $F_V=2G_V$ in the
anti-symmetric tensor formalism leads to
the appearance of the non-minimal $\rho\pi\pi$ coupling $z_4$.
This coupling actually violates the physical KSRF I relation, while
it does not contribute to the $\rho\pi\pi$ coupling at zero
momentum keeping the low energy theorem of the hidden local
symmetry\cite{kn:BKY85,kn:HY92}.

We next discuss the axial-vector meson (the $a_1$ meson) in the chiral
lagrangian.
In the anti-symmetric tensor field method, it is straightforward to
introduce the $a_1$ meson\cite{kn:EGPR89,kn:EGLPR89}:
\begin{eqnarray}
  {\cal L}_{\rm AST}
    &=& \frac{f_\pi^2}{4} \mbox{tr}((D^\mu U)^\dagger (D_\mu U))
    \nonumber\\
    & & -\frac{1}{2} \mbox{tr}(\nabla^\lambda {\bf A}_{\lambda\mu}
                               \nabla_\nu {\bf A}^{\nu\mu})
        +\frac{M_A^2}{4} \mbox{tr}({\bf A}_{\mu\nu}{\bf A}^{\mu\nu})
        -\frac{F_A}{\sqrt{2}}\tr({\bf A}^{\mu\nu} \hat{\cal A}_{\mu\nu}),
    \nonumber\\
    & & -\frac{1}{2} \mbox{tr}(\nabla^\lambda {\bf V}_{\lambda\mu}
                               \nabla_\nu {\bf V}^{\nu\mu})
        +\frac{M_V^2}{4} \mbox{tr}({\bf V}_{\mu\nu}{\bf V}^{\mu\nu})
        +\frac{F_V}{\sqrt{2}}\tr({\bf V}^{\mu\nu} \hat{\cal V}_{\mu\nu})
    \nonumber\\
    & & +\frac{G_V}{2\sqrt{2}}
           i \mbox{tr}({\bf V}^{\mu\nu}[u_\mu,u_\nu])
\label{eq:ast2}
\end{eqnarray}
where $M_A$ and $F_A$ are the mass and the decay constant of the $a_1$
meson respectively, and $\hat{\cal A}_{\mu\nu}$ is defined by
\begin{displaymath}
\hat{\cal A}_{\mu\nu}
  = \frac{1}{2}
         ( -\xi_L {\cal L}_{\mu\nu}\xi_L^\dagger
           +\xi_R {\cal R}_{\mu\nu}\xi_R^\dagger ).
\end{displaymath}
The anti-symmetric tensor field ${\bf A}_{\mu\nu}$ represents
the $a_1$ meson.

Bando, Kugo and Yamawaki (BKY) introduced the $a_1$ meson as a gauge
field of generalized hidden local symmetry ($H_L\times
H_R=SU(N)\times SU(N)$)\cite{kn:BKY85n}:
\begin{equation}
  U=\xi_L^\dagger \xi_M \xi_R,
\end{equation}
where $\xi_L,\xi_R,\xi_M$ are transforming as
\begin{displaymath}
  \xi_L \rightarrow h_L \xi_L g_L^\dagger, \quad
  \xi_R \rightarrow h_R \xi_L g_R^\dagger, \quad
  \xi_M \rightarrow h_L \xi_M h_R^\dagger, \quad
  h_L \in H_L, \quad h_R \in H_R.
\end{displaymath}
Corresponding to this symmetry, BKY introduced gauge fields $L_\mu$
and $R_\mu$:
\begin{displaymath}
  D_\mu \xi_L = \partial\xi_L -iL_\mu \xi_L + i\xi_L {\cal L}_\mu,
  \quad
  D_\mu \xi_R = \partial\xi_R -iR_\mu \xi_R + i\xi_R {\cal R}_\mu,
  \quad
  D_\mu \xi_M = \partial\xi_M -iL_\mu \xi_M + i\xi_M R_\mu.
\end{displaymath}
The BKY lagrangian is written as:
\begin{eqnarray}
  {\cal L}_{\rm BKY}
  &=& a f_\pi^2 \mbox{tr}\left(
          \hat \beta_{\mu\parallel}\xi_M^\dagger
          \hat \beta^\mu_\parallel \xi_M^\dagger
      \right)
     +b f_\pi^2 \mbox{tr}\left(
          \left( \hat \beta_{\mu\perp}\xi_M^\dagger
                +\frac{1}{2}\hat\beta_{\mu M} \right)^2
      \right)
     +\frac{c}{4} f_\pi^2 \mbox{tr}\left(
          \hat\beta_{\mu M} \hat\beta^\mu_M
      \right)
  \nonumber\\
  & &
     +d f_\pi^2 \mbox{tr} \left(
          \hat \beta_{\mu\perp}\xi_M^\dagger
          \hat \beta^\mu_\perp \xi_M^\dagger
      \right)
     -\frac{1}{4g^2} \left[
         \mbox{tr}\left(L_{\mu\nu}L^{\mu\nu}\right)
        +\mbox{tr}\left(R_{\mu\nu}R^{\mu\nu}\right)
      \right],
\label{eq:bky}
\end{eqnarray}
where $\hat \beta_{\mu\parallel}$,$\hat \beta_{\mu\perp}$ and $\hat
\beta_{\mu M}$ are given by
\begin{displaymath}
  \hat\beta_{\mu\parallel} \equiv
    \frac{1}{2} (\hat\beta_{\mu L} \xi_M + \xi_M \hat\beta_{\mu R}),
  \qquad
  \hat\beta_{\mu\perp} \equiv
    \frac{1}{2} (\hat\beta_{\mu L} \xi_M - \xi_M \hat\beta_{\mu R}
                                  - \hat\beta_{\mu M}\xi_M),
\end{displaymath}
and
\begin{displaymath}
  \hat\beta_{\mu L}\equiv \frac{1}{i} D_\mu \xi_L \cdot \xi_L^\dagger,
  \quad
  \hat\beta_{\mu R}\equiv \frac{1}{i} D_\mu \xi_R \cdot \xi_R^\dagger,
  \quad
  \hat\beta_{\mu M}\equiv \frac{1}{i} D_\mu \xi_M \cdot \xi_M^\dagger.
\end{displaymath}
The linear combination $L_\mu \pm R_\mu$ correspond to $\rho$ and $a_1$
mesons respectively.

To show the equivalence of the two formulations, we first rewrite
the lagrangian (\ref{eq:ast2}) so as to make the generalized hidden
local symmetry manifest:
\begin{eqnarray}
  {\cal L}_{\rm AST}
    &=& \frac{f_\pi^2}{4} \mbox{tr}((D^\mu U)^\dagger (D_\mu U))
    \nonumber\\
    & & -\frac{1}{2} \mbox{tr}(\nabla^\lambda {\bf A}_{\lambda\mu}
                               \xi_M^\dagger
                               \nabla_\nu {\bf A}^{\nu\mu}\xi_M^\dagger)
        +\frac{M_A^2}{4} \mbox{tr}({\bf A}_{\mu\nu}\xi_M^\dagger
                                   {\bf A}^{\mu\nu}\xi_M^\dagger)
        -\frac{F_A}{\sqrt{2}}\tr({\bf A}^{\mu\nu} \xi_M^\dagger
                                 \hat{\cal A}_{\mu\nu}\xi_M^\dagger),
    \nonumber\\
    & & -\frac{1}{2} \mbox{tr}(\nabla^\lambda {\bf V}_{\lambda\mu}
                               \xi_M^\dagger
                               \nabla_\nu {\bf V}^{\nu\mu}
                               \xi_M^\dagger)
        +\frac{M_V^2}{4} \mbox{tr}({\bf V}_{\mu\nu}\xi_M^\dagger
                                   {\bf V}^{\mu\nu}\xi_M^\dagger)
        +\frac{F_V}{\sqrt{2}}\tr({\bf V}^{\mu\nu} \xi_M^\dagger
                                  \hat{\cal V}_{\mu\nu} \xi_M^\dagger)
    \nonumber\\
    & & +\frac{G_V}{2\sqrt{2}}
           i \mbox{tr}({\bf V}^{\mu\nu} \xi_M^\dagger
                       [u_\mu \xi_M^\dagger, u_\nu \xi_M^\dagger] ),
\label{eq:ast3}
\end{eqnarray}
where the covariant derivative $\nabla_\mu$ and the chiral covariant one
form $u_\mu$ are redefined as:
\begin{displaymath}
  \nabla^\mu {\bf V}_{\mu\nu}
    = \partial^\mu {\bf V}_{\mu\nu} + \Gamma^\mu_L {\bf V}_{\mu\nu}
                                    - {\bf V}_{\mu\nu} \Gamma^\mu_R,
  \qquad
  \nabla^\mu {\bf A}_{\mu\nu}
    = \partial^\mu {\bf A}_{\mu\nu} + \Gamma^\mu_L {\bf A}_{\mu\nu}
                                    - {\bf A}_{\mu\nu} \Gamma^\mu_R,
\end{displaymath}
and
\begin{displaymath}
  u_\mu = i \xi_L (D_\mu U) \xi_R^\dagger
\end{displaymath}
with
\begin{eqnarray*}
  \Gamma_L^\mu
    &=& -\frac{1}{2}\left[
          \partial^\mu \xi_L \cdot \xi_L^\dagger
         +\partial^\mu (\xi_M\xi_R) \cdot \xi_R^\dagger\xi_M^\dagger
         +i\xi_L {\cal L}^\mu \xi_L^\dagger
         +i\xi_M\xi_R {\cal R}^\mu \xi_R^\dagger \xi_M^\dagger
         \right],
    \\
  \Gamma_R^\mu
    &=& -\frac{1}{2}\left[
          \partial^\mu \xi_R \cdot \xi_R^\dagger
         +\partial^\mu (\xi_M^\dagger\xi_L) \cdot \xi_L^\dagger\xi_M
         +i\xi_R {\cal R}^\mu \xi_R^\dagger
         +i\xi_M^\dagger\xi_L {\cal L}^\mu \xi_L^\dagger \xi_M
         \right].
\end{eqnarray*}
The external fields $\hat{\cal V}_{\mu\nu}$ and $\hat{\cal
A}_{\mu\nu}$ are redefined as
\begin{displaymath}
  \hat{\cal V}_{\mu\nu}
  = \frac{1}{2}
         ( \xi_L {\cal L}_{\mu\nu}\xi_L^\dagger \xi_M
          +\xi_M \xi_R {\cal R}_{\mu\nu}\xi_R^\dagger ),
  \quad
  \hat{\cal A}_{\mu\nu}
  = \frac{1}{2}
         ( -\xi_L {\cal L}_{\mu\nu}\xi_L^\dagger \xi_M
           +\xi_M \xi_R {\cal R}_{\mu\nu}\xi_R^\dagger ).
\end{displaymath}
The ``matter'' fields ${\bf V}_{\mu\nu}$ and ${\bf A}_{\mu\nu}$
transform as
\begin{displaymath}
  {\bf V}_{\mu\nu} \rightarrow  h_L {\bf V}_{\mu\nu} h_R^\dagger,
 \qquad
  {\bf A}_{\mu\nu} \rightarrow  h_L {\bf A}_{\mu\nu} h_R^\dagger,
\end{displaymath}
in the lagrangian (\ref{eq:ast3}).
It is easy to see that (\ref{eq:ast3}) actually reproduces its
original form
(\ref{eq:ast2}) in the unitary gauge of the generalized hidden local
symmetry, i.e., $\xi_M=1$ and $\xi_L^\dagger = \xi_R$.
We also obtain following relations:
\begin{eqnarray*}
  \Gamma_L^\mu
    &=& -\frac{i}{2} (
          \hat\beta_L^\mu + \xi_M \hat\beta_R^\mu \xi_M^\dagger
         +\hat\beta_M^\mu + 2L_\mu ),
    \\
  \Gamma_R^\mu
    &=& -\frac{i}{2} (
          \hat\beta_R^\mu + \xi_M^\dagger \hat\beta_L^\mu \xi_M
         -\xi_M^\dagger \hat\beta_M^\mu \xi_M + 2R_\mu ),
    \\
  u_\mu
    &=& 2\hat\beta_{\mu\perp}.
\end{eqnarray*}

The equivalence of the lagrangian (\ref{eq:ast3}) and the BKY
formalism (\ref{eq:bky}) can be shown in a similar manner to the case
of the $\rho$ meson.
We introduce spin 1 fields $L_\mu$ and $R_\mu$ as auxiliary fields:
\begin{eqnarray}
  {\cal L}_{\rm AST}'
  &=& {\cal L}_{\rm AST}
  \nonumber\\
  & &+ \frac{\kappa^2}{4}\mbox{tr}\left[
         \left( L_\mu +i\partial_\mu\xi_L\cdot \xi_L^\dagger
                      - \xi_L {\cal L}_\mu \xi_L^\dagger
               -\frac{1}{\kappa}\nabla^\lambda {\bf V}_{\lambda\mu}
                        \cdot \xi_M^\dagger
               +\frac{1}{\kappa}\nabla^\lambda {\bf A}_{\lambda\mu}
                        \cdot \xi_M^\dagger
         \right)^2\right]
  \nonumber\\
  & &+ \frac{\kappa^2}{4}\mbox{tr}\left[
         \left( R_\mu +i\partial_\mu\xi_R\cdot \xi_R^\dagger
                      - \xi_R {\cal R}_\mu \xi_R^\dagger
               -\frac{1}{\kappa}\xi_M^\dagger \cdot
                         \nabla^\lambda {\bf V}_{\lambda\mu}
               -\frac{1}{\kappa}\xi_M^\dagger\cdot
                         \nabla^\lambda {\bf A}_{\lambda\mu}
         \right)^2\right],
\label{eq:aux}
\end{eqnarray}
with $\kappa$ being parameters corresponding the arbitrariness of the
definition of the effective spin 1 meson fields.
By performing partial integration and using the identities
\begin{eqnarray*}
  D_\mu \hat\beta_{\nu L} - D_\nu \hat\beta_{\mu L}
  &=& -L_{\mu\nu} + \xi_L {\cal L}_{\mu\nu} \xi_L^\dagger
                  +i[\hat\beta_{\mu L}, \hat\beta_{\nu L}],
  \\
  D_\mu \hat\beta_{\nu R} - D_\nu \hat\beta_{\mu R}
  &=& -R_{\mu\nu} + \xi_R {\cal R}_{\mu\nu} \xi_R^\dagger
                  +i[\hat\beta_{\mu R}, \hat\beta_{\nu R}],
  \\
  D_\mu \hat\beta_{\nu M} - D_\nu \hat\beta_{\mu M}
  &=& -L_{\mu\nu} + \xi_M R_{\mu\nu} \xi_R^\dagger
                  +i[\hat\beta_{\mu M}, \hat\beta_{\nu M}],
\end{eqnarray*}
we find (\ref{eq:aux}) can be rewritten as
\begin{eqnarray}
  {\cal L}'_{\rm AST}
  &=& \frac{\kappa^2}{4}\mbox{tr}(\hat\beta_{\mu L}\hat\beta^\mu_L)
     +\frac{\kappa^2}{4}\mbox{tr}(\hat\beta_{\mu R}\hat\beta^\mu_R)
     +f_\pi^2 \mbox{tr}(\hat\beta_{\mu\perp}\xi_M^\dagger
                        \hat\beta^\mu_\perp \xi_M^\dagger)
  \nonumber\\
  & & +\frac{M_V^2}{4}\mbox{tr}\left[
         ({\bf V}_{\mu\nu}\xi_M^\dagger+X_{V\mu\nu}\xi_M^\dagger)^2
       \right]
      +\frac{M_A^2}{4}\mbox{tr}\left[
         ({\bf A}_{\mu\nu}\xi_M^\dagger+X_{A\mu\nu}\xi_M^\dagger)^2
       \right]
  \nonumber\\
  & & -\frac{M_V^2}{4}\mbox{tr}\left[
         X_{V\mu\nu}\xi_M^\dagger X_V^{\mu\nu} \xi_M^\dagger
       \right]
       -\frac{M_A^2}{4}\mbox{tr}\left[
         X_{A\mu\nu}\xi_M^\dagger X_A^{\mu\nu} \xi_M^\dagger
       \right],
\label{eq:aux2}
\end{eqnarray}
where $X_V$ and $X_A$ are defined by
\begin{eqnarray*}
  X_V^{\mu\nu}
  &\equiv&
    \sqrt{2} F_V \hat{\cal V}^{\mu\nu}
  + 2\sqrt{2}G_V i[\hat\beta^\mu_\perp\xi_M^\dagger,
                   \hat\beta^\nu_\perp\xi_M^\dagger]\xi_M
  -\frac{\kappa}{2}\left(
       2\hat{\cal V}^{\mu\nu} - L^{\mu\nu}\xi_M -\xi_M R^{\mu\nu}
      \right)
  \nonumber\\
  & & +\frac{\kappa}{2}\left(
       i[\hat\beta^\mu_L, \xi_M \hat\beta^\nu_R \xi_M^\dagger]\xi_M
       +\frac{i}{2} [\hat\beta^\mu_M,
           \hat\beta^\nu_L-\xi_M\hat\beta^\nu_R\xi_M^\dagger]\xi_M
       - (\mu \leftrightarrow \nu)\right),
  \\
  X_A^{\mu\nu}
  &\equiv&
    -\sqrt{2} F_A \hat{\cal A}^{\mu\nu}
    +\frac{\kappa}{2}\left(
       2\hat{\cal A}^{\mu\nu} + L^{\mu\nu}\xi_M -\xi_M R^{\mu\nu}
      \right)
  \nonumber\\
  & & +\frac{\kappa}{2}\left(
       \frac{i}{2} [\hat\beta^\mu_M,
           \hat\beta^\nu_L+\xi_M\hat\beta^\nu_R\xi_M^\dagger]\xi_M
       - (\mu \leftrightarrow \nu)\right),
\end{eqnarray*}
It is easy to integrate out the anti-symmetric tensor fields
${\bf V}_{\mu\nu}$ and ${\bf A}_{\mu\nu}$ from (\ref{eq:aux2}).
We obtain
\begin{eqnarray}
  {\cal L}'_{\rm AST}
  &=& \frac{\kappa^2}{4}\mbox{tr}(\hat\beta_{\mu L}\hat\beta^\mu_L)
     +\frac{\kappa^2}{4}\mbox{tr}(\hat\beta_{\mu R}\hat\beta^\mu_R)
     +f_\pi^2 \mbox{tr}(\hat\beta_{\mu\perp}\xi_M^\dagger
                        \hat\beta^\mu_\perp \xi_M^\dagger)
  \nonumber\\
  & & -\frac{M_V^2}{4}\mbox{tr}\left[
         X_{V\mu\nu}\xi_M^\dagger X_V^{\mu\nu} \xi_M^\dagger
       \right]
       -\frac{M_A^2}{4}\mbox{tr}\left[
         X_{A\mu\nu}\xi_M^\dagger X_A^{\mu\nu} \xi_M^\dagger
       \right].
\label{eq:aux3}
\end{eqnarray}
The expression (\ref{eq:aux3}) is invariant under generalized
hidden local symmetry and includes the spin 1 fields as gauge fields.
The lagrangian (\ref{eq:aux3}) includes left-right gauge field
kinetic mixing term $\mbox{tr}(L_{\mu\nu}\xi_M R^{\mu\nu} \xi_M^\dagger)$,
which is not included in the BKY formalism, however.

We have three ambiguities of the definition of the effective fields
$L_\mu$ and $R_\mu$ corresponding to the three chiral covariant one
forms $\hat\beta_{\mu L}$, $\hat\beta_{\mu R}$ and $\hat\beta_{\mu M}$.
One ambiguity is already taken into account by the parameter $\kappa$.
The following redefinition is convenient to parametrize the rest of
the ambiguities:
\begin{eqnarray*}
  L_\mu &\rightarrow&
    L_\mu + \lambda\mu \hat\beta_{\mu L}
   -(\lambda-\mu)\xi_M \hat\beta_{\mu R} \xi_M^\dagger
   +(1-\lambda)\mu \hat\beta_{\mu M},
  \\
  R_\mu &\rightarrow&
    R_\mu + \lambda\mu \hat\beta_{\mu R}
   -(\lambda-\mu)\xi_M^\dagger \hat\beta_{\mu L} \xi_M
   -(1-\lambda)\mu \xi_M^\dagger\hat\beta_{\mu M}\xi_M,
\end{eqnarray*}
where $\lambda$ and $\mu$ are arbitrary parameters.
We can eliminate the above mentioned term $\mbox{tr}(L_{\mu\nu}\xi_M
R^{\mu\nu} \xi_M^\dagger)$ by using the parameter $\lambda$.
The parameter $\mu$ represents the rest of the ambiguity.

A plausible choice of the parameters $\kappa$,$\lambda$ and $\mu$ is
to determine them
so as to eliminate the gauge kinetic mixing terms, e.g,
$\mbox{tr}(L_{\mu\nu}\xi_L {\cal L}^{\mu\nu}\xi_L^\dagger)$ in the
effective lagrangian.
This particular choice of effective field definition leads to
\begin{equation}
  \kappa = -\frac{M_A-M_V}{2\sqrt{2}}
             \left(\frac{F_V}{M_V} + \frac{F_A}{M_A}\right),
         \quad
  \lambda= -\frac{M_A+M_V}{M_A-M_V},
         \quad
  \mu    = -\frac{F_V M_A - F_A M_V}{F_V M_A + F_A M_V}.
\end{equation}
The lagrangian (\ref{eq:aux3}) then becomes
\begin{eqnarray}
  {\cal L}_{\rm AST}'
  &=& -\frac{F_V^2}{4M_V^2} \left[
         \mbox{tr}(L_{\mu\nu}L^{\mu\nu})+\mbox{tr}(R_{\mu\nu}R^{\mu\nu})
       \right]
  \nonumber\\
  & &+F_V^2 \mbox{tr}\left(
        \hat\beta_{\mu\parallel}\xi_M^\dagger
        \hat\beta^\mu_\parallel \xi_M^\dagger
      \right)
     +F_V F_A \frac{M_A}{M_V} \mbox{tr}\left(
          \left( \hat \beta_{\mu\perp}\xi_M^\dagger
                +\frac{1}{2}\hat\beta_{\mu M} \right)^2
      \right)
  \nonumber\\
  & &
     +\frac{1}{4} \left[
       F_V^2 \frac{M_A^2}{M_V^2} - F_V F_A \frac{M_A}{M_V}
      \right]
      \mbox{tr}\left(
          \hat\beta_{\mu M} \hat\beta^\mu_M
      \right)
     +\left[
        f_\pi^2 + F_A^2 - F_V F_A \frac{M_A}{M_V}
      \right]
      \mbox{tr} \left(
          \hat \beta_{\mu\perp}\xi_M^\dagger
          \hat \beta^\mu_\perp \xi_M^\dagger
      \right)
   \nonumber\\
   & &  +\cdots,
\end{eqnarray}
where $\cdots$ stands for higher derivative terms in the hidden local
symmetry formulation.  We thus obtain explicit relations between the
anti-symmetric tensor method and the generalized hidden local symmetry
formalism:
\vspace*{-1cm}
\begin{equation}
\begin{minipage}[t]{3in}
\begin{eqnarray*}
  g &=& \frac{M_V}{F_V}, \\
  b &=& \frac{F_V F_A }{f_\pi^2} \frac{M_A}{M_V}, \\
  d &=& 1 + \frac{F_A^2}{f_\pi^2}
          - \frac{F_V F_A}{f_\pi^2} \frac{M_A}{M_V}.
\end{eqnarray*}
\end{minipage}
\begin{minipage}[t]{3in}
\begin{eqnarray*}
  a &=& \frac{F_V^2}{f_\pi^2}, \\
  c &=& \frac{F_V^2 M_A^2}{f_\pi^2 M_V^2}
       -\frac{F_V F_A M_A}{f_\pi^2 M_V},
\end{eqnarray*}
\end{minipage}
\label{eq:abcd}
\end{equation}
Plugging the phenomenological value $F_V^2 \simeq 2 f_\pi^2$ and the
Weinberg sum rules $f_\pi^2=F_V^2 -F_A^2$ and $F_V^2 M_V^2 = F_A^2
M_A^2$ in (\ref{eq:abcd}) , we obtain the coefficients $a=b=c=2$, $d=0$,
in agreement with the values quoted in Ref.\cite{kn:BKY88}.

In this paper, we have shown how the auxiliary field method works to
clarify the relation
of the anti-symmetric tensor and the hidden local symmetry formalisms
of the $\rho$ and the $a_1$ mesons.
The ambiguity of the definition of the effective fields in the hidden
local symmetry formalism can be resolved by using the extra
conditions, e.g., the disappearance of the kinetic mixing terms in the
effective lagrangian.
The anti-symmetric tensor field method is equivalent to the hidden
local symmetry lagrangian plus on-shell KSRF I violating
${\cal O}(E^4)$ term.
For analysis of non QCD-like technicolor models, this term might
become important.
\vspace{0.3cm}

The author thanks M. S. Chanowitz, Y. Okada, M. Suzuki and K. Yamawaki
for enlightening discussions. He is also grateful to B. Bullock for
careful reading of the manuscript.

\begin{flushleft}
  {\em Note added}:
\end{flushleft}
After the manuscript has been completed, I realized similar work of
J. Bijnens and E. Pallante\cite{kn:BP95}.

\end{document}